# Transient Synchronization Stability Analysis of Wind Farms with MMC-HVDC Integration Under Offshore AC Grid Fault

Yu Zhang[1], Chen Zhang[1], Renxin Yang[1], Jing Lyu[1], Li Liu[1,2], Xu Cai[1*]

[1] Key Laboratory of Control of Power Transmission and Conversion, Shanghai Jiao Tong University, Ministry of Education, Minhang District, Shanghai, China;
[2] Zhoushan Power Supply Company, National Grid Zhejiang Electric Power Corporation, Zhoushan, Zhejiang, China;
*xucai@sjtu.edu.cn

**Abstract:** The MMC-HVDC connected offshore wind farms (OWFs) could suffer short circuit fault (SCF), whereas their transient stability is not well analysed. In this paper, the mechanism of the loss of synchronization (LOS) of this system is analysed considering the whole system state from the fault-on to the post-fault, and the discussion on fault type and fault clearance is addressed as well. A stability index is proposed to quantify the transient synchronization stability (TSS) of the system, which is capable to not only estimate whether the wind turbine generators (WTGs) be able to get resynchronized with the offshore MMC after the fault is cleared, but also to evaluate the performance of stability improving methods as well. Finally, a scenario of six cases is tested on the PSCAD/EMTDC simulation platform, where the performances of four existing stability improving methods are thoroughly compared via both numerical simulation and the proposed stability index.

## 1 Introduction

Nowadays, renewable energy has been widely utilized for power generation. An increasing amount of offshore wind farms (OWFs) has been built in order to exploit the premium wind resource from the far sea. For OWFs over 80~120km from the coastline, HVDC transmission system is more qualified than the HVAC, thanks to the non-necessity of reactive power compensation. So far, the most prevailing HVDC system for OWF grid-integration is based on the modular multi-level converter (MMC).

In the past decade, it is found that the OWF-MMC system would suffer instability problems due to the interaction between OWFs and MMC station [1, 2], likely tripping the wind turbines when the interaction becomes severe. However, the previous research mainly focused on the small-signal stability near a given working point, therefore, stability analysis of this scope can be readily approached by linear control theorem such as the Nyquist criterion [2]. In contrast, when the OWF-MMC system is subjected to large disturbance, stability issues of this scope might be more difficult to analyse because of lacking generic large-signal stability analysis tools, and requiring the knowledge of the system's fault ride-through (FRT) strategies.

The FRT strategy during the short circuit fault (SCF) of the sending end AC grid is very challenging, because overcurrent is likely to occur in the sending end MMC station, as the controlling scope is to maintain the offshore AC grid voltage. Several current limiting controls were put forward to supress the fault current [3-5]. The common goal of these methods is to reduce the output voltage by measuring the amplitude of either the AC side grid voltage or current, which is not very robust, and time delays may be introduced. To this end, the dual-loop control structure [6] (voltage control with inner current loop) is opted for offshore MMC station control because the inner loop could effectively confine the output current to the set-limit. The implementation of the outer loop voltage control with MMC is challenging for its AC side does not have a physical capacitance [7], yet the dual loop scheme is so far the optimal solution for offshore MMC control. Besides, [8-10] solved asymmetric fault ride through with dual-loop control, transient overvoltage and suppression of DC short circuit current issues. In general, the FRT capability of OWF-MMC system has been greatly improved.

Aside from the analysis of FRT strategies, recent finding has shown that OWFs could suffer loss of synchronization (LOS) issue under offshore AC grid fault [11]. The root cause is related to the nonlinear dynamics of the phase-locked loop (PLL) employed in the grid interfacing converter control of the WT. One of the most evident consequence of the LOS is that the wind turbine can no longer get synchronized with the grid even if the fault is cleared. This issue is often referred to as the transient synchronization stability (TSS) [12-17] in the context of wind power integration. Current analysis of the TSS issue is mainly focused on the existence of equilibrium point (EP) during the high-impedance grid fault and whether the existent EP is stable. However, most of these analyses are based on idealized system while many practical issues which are of great significance to TSS analysis have not been properly considered and discussed, e.g., the effect of FRT control, the relay protection of the cable line, and the resynchronization process of the PLL, etc.

This paper aims to fill the above gaps by providing a more detailed TSS analysis of the OWF-MMC system. In this paper, the LOS mechanism of the OWF-MMC system is analysed considering the whole process of FRT (e.g., the FRT control intervene and fault clearance of the grid); to quantify the impact of TSS, a stability index is proposed; and by using the stability index, the performances of some introduced stability enhancement methods are compared and verified by time-domain simulations in PSCAD/EMTDC.

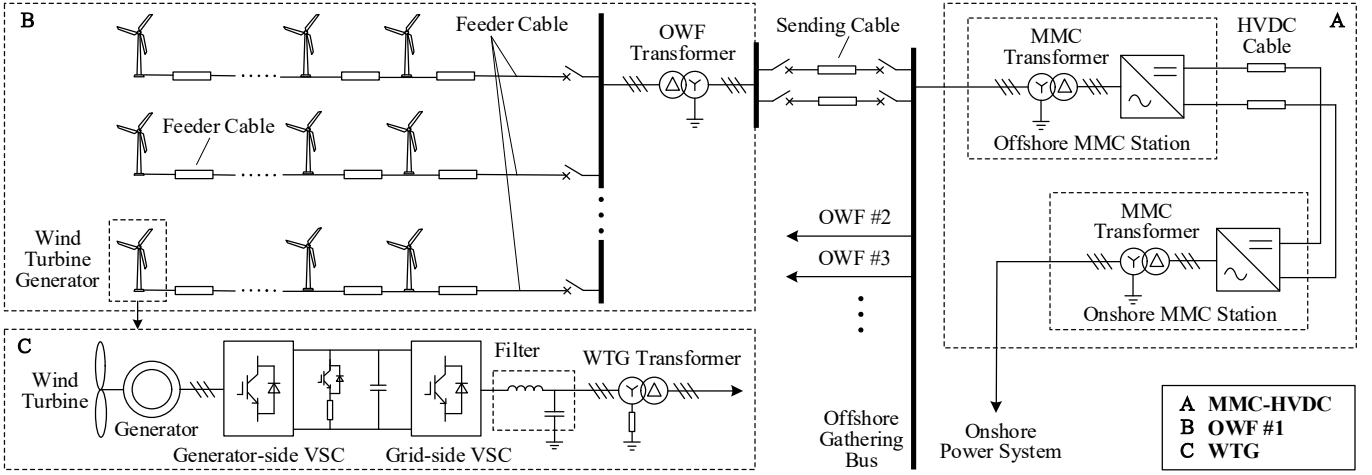

Fig. 1 Configuration of OWF-MMC system.

## 2 Modelling of OWF-MMC system

### 2.1 System Configuration

A typical configuration of OWF MMC-HVDC system is illustrated in Fig. 1. The MMC-HVDC system is connected with several OWFs. For each OFW, there are usually tens of feeder lines each with 4~8 wind turbine generators (WTGs) connection. Type-4 wind turbine generators (WTGs) are most likely to be installed for offshore wind power generation. The voltage is stepped twice from the WTG to the offshore MMC station. For a specific project case, the voltage is stepped up from 690V to 35kV by the WTG transformer, and from 35kV to 220kV by the OWF transformer. To ensure the reliability of each OWF, double circuit sending cables are laid to connect the OWF with the offshore MMC station, and AC circuit breakers (CBs) are equipped at each end.

### 2.2 Control and FRT strategy of the OWF-MMC System

The voltage of the offshore AC system is formed by the offshore MMC station via $V/f$ control, as is shown in Fig. 2. Dual-loop controller with negative sequence structure is used to control the voltage of offshore gathering bus (OGB). The positive and negative sequence voltage and current are decomposed by the double second-order generalized integrator (DSOGI) [18]. Positive sequence $d$-axis voltage reference is set to 1 p.u., and the frequency is constant 50Hz. The elliptic current limiter is equipped to limit the amplitude of phase current [19]. To ride through both symmetric and asymmetric SCF in the offshore AC grid, a modification to the positive $d$-axis voltage reference and an individual negative sequence current limiter are introduced [8], thus the offshore MMC station could still establish the AC voltage well during SCF without fault detection and controller switching.

With regard to the WTG, the control structure is shown in Fig. 3. The grid side VSC is operating in $U_{dc}/Q$ mode. The negative sequence current controller is also installed to supress the unbalanced current during the asymmetric SCF. The generator-side VSC usually performs the maximum power point tracking (MPPT) control. During the occurrence of the SCF, the positive sequence voltage of the AC grid would drop, and therefore the WTG must switch to the low-voltage right through (LVRT) control. The chopper circuit would be activated when the dc voltage exceeds the threshold. Besides, certain amount of dynamic reactive current should be injected during the SCF depending on the voltage amplitude. For example, according to the grid code of China, the $q$-axis current injected to the grid during the LVRT period should be:

$$I_{qref+} = -1.5 \times (0.9 - U_+) . \quad (1)$$

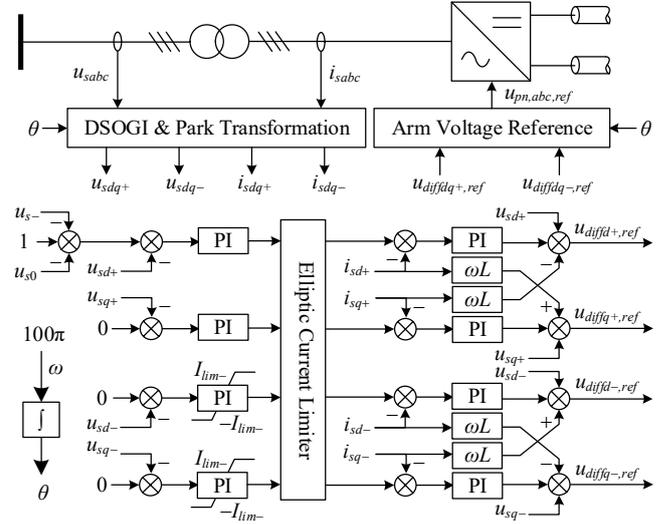

Fig. 2 Control of offshore MMC station.

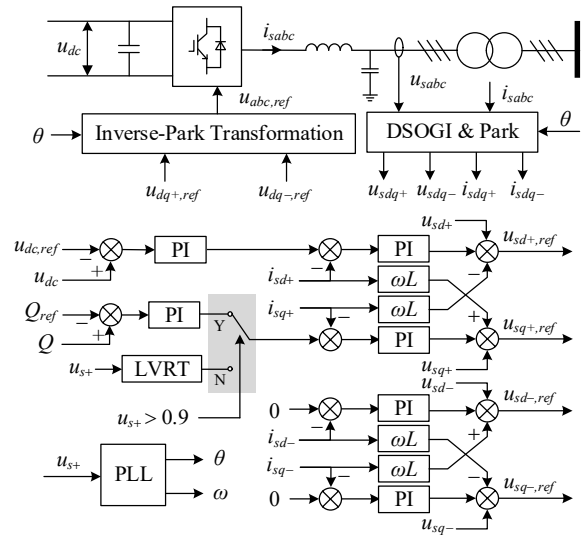

Fig. 3 Control of the grid side VSC of WTG.

Besides, $d$ axis current would possibly decrease because of the maximum current limitation and the priority in $q$ axis current during the LVRT stage.

*2.3 Simplified Modelling for TSS Analysis*

To analyse the TSS issue of the OWF-MMC system, a simplified modelling is derived in the below.

First, all the WTGs in OWF are aggregated as a single to represent the whole, and the single WTG could be further represented by a controlled current source [14], as is shown in Fig. 4. Its current is regarded equal to the reference computed by the outer-loop controller, and the phase angle $\theta$ is generated from the SRF-PLL [20], as is illustrated in Fig. 5.

Next, the feeder cables inside the OWF are aggregated via the method proposed in [21], thus the grid impedance $Z_g$ is composed of four parts: 1) the WTG transformer leakage impedance, 2) the aggregated feeder cable, 3) the OWF transformer, and 4) the 220kV sending cable. If the SCF occurs on the offshore AC grid or any CBs takes action, then $Z_g$ would be changed accordingly. It should be noted that this system is still inductance-dominant because the cables in this system are not long (from hundreds of meters to ten kilometres), thus the admittance of the offshore transmission are ignorable.

Finally, offshore MMC station is modelled as a voltage source with constant frequency according to its control scope. Different from ordinary voltage source, during the SCF in the offshore AC grid, the MMC could possibly work in three modes automatically [11], depending on the fault type and fault location, and either overmodulation or current limiting mode would occur if the fault is severe. In general, however, regardless of the working mode during the fault, the offshore MMC is equivalent to a voltage source, yet the voltage amplitude could be lower than the rated value.

Based on the above elaboration, the OFF-MMC system could be represented by the simplified circuit in Fig. 4, where a controlled current source is connected to a voltage source through an impedance, which is the most commonly used model for TSS analysis in the previous study [14, 16]. In this model, the voltage of the point of connection (POC) is:

$$U_s = U_g + Z_g I_s , \quad (2)$$

where $U_g = U_g \cdot e^{j\theta_g}$ is the voltage phasor of the offshore MMC station ($\theta_g = \omega_s t$, $\omega_s$ is the synchronous radius frequency), $Z_g = R_g + j(\omega_{pll}/\omega_s) X_g$ is the equivalent grid impedance ($\omega_{pll}$ is the instantaneous frequency of PLL), and $I_s = (I_{sd} + jI_{sq}) \cdot e^{j\theta}$ is the output current of the aggregated WTG ($\theta$ is the angle of PLL). Then, the $q$-axis POC voltage

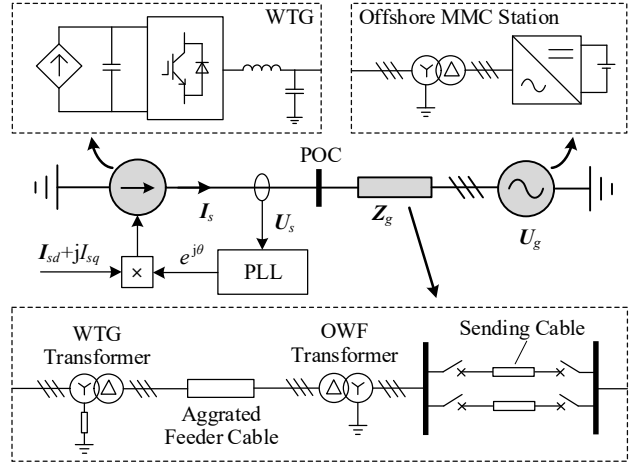

Fig. 4 Simplified model of OWF-MMC system for TSS analysis.

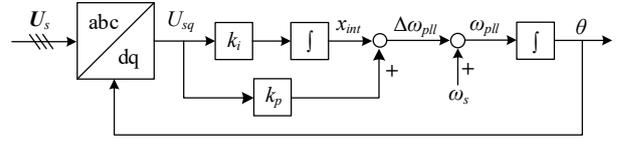

Fig. 5 Block diagram of a typical SRF-PLL.

could be derived by multiplying $e^{-j\theta}$ to each side of (2) and keep the imaginary part:

$$U_{sq} = -U_g \sin\delta + R_g I_{sq} + \omega_{pll} X_g I_{sd}/\omega_s , \quad (3)$$

where $\delta = \theta - \theta_{mmc}$ is the angle difference between PLL and the voltage phasor of the MMC which is also equivalent to the power angle of POC in steady state.

Next, according to the block diagram in Fig. 5, the SRF-PLL obeys the following equations:

$$\begin{cases} \dot{\delta} = x_{int} + k_p U_{sq} \\ \dot{x}_{int} = k_i U_{sq} \end{cases}, \quad (4)$$

According to (3) and (4), the dynamic equations of the PLL system could be obtained, which is:

$$\begin{cases} \dot{\delta} = \dfrac{k_p(U_0 - U_g \sin\delta) + x_{int}}{1 - k_p X_g I_{sd}/\omega_s} \\ \dot{x}_{int} = \dfrac{k_i(U_0 - U_g \sin\delta) + k_i X_g I_{sd}/\omega_s \cdot x_{int}}{1 - k_p X_g I_{sd}/\omega_s} \end{cases}, \quad (5)$$

where $U_0 = R_g I_{sq} + X_g I_{sd}$ denotes the voltage offset.

This set of equations governs the synchronizing behaviour of the OWF-MMC system, which is influenced by not only the system parameters, but also the variation of grid connection (this would change the impedance), LVRT strategy of the WTG (current reference), working mode of offshore MMC station (grid voltage). In addition, the fault clearing time (FCT) also affects the dynamic behaviour of the system, which will be revealed in the following part.

## 3 TSS Analysis of OWF-MMC System

*3.1 Mechanism of the LOS in OWF-MMC system*
*3.1.1 Current injection boundary:*

According to (5), the output current should be bounded, so that the system has steady state point, where the following

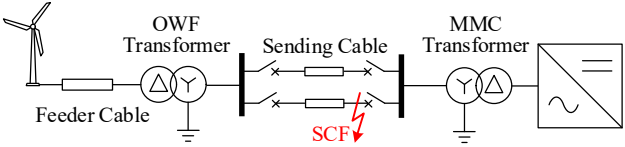

Fig. 6 Illustration of SCF of the OWF-MMC system.

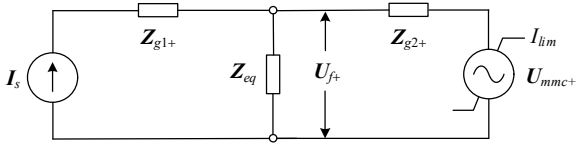

Fig. 7 Positive sequence circuit when SCF occurs.

condition should be satisfied:

$$|R_g I_{sq} + X_g I_{sd}| < U_g \quad (6)$$

This condition is referred to as the current injection boundary. Generally, the grid resistance is relatively small compared to the reactance in this cable system, so (6) could be reduced to:

$$|X_g I_{sd}| < U_g \quad (7)$$

In normal operation, the output current is always within the boundary, whereas it might be violated during SCF due to very low remaining voltage and much $d$-axis current injection to the inductance-dominant grid. Consequently, if (6) is not satisfied, both $\delta$ and $x_{int}$ will continuously increase, leading to separation of the power angle between WTG and MMC, and the output frequency of PLL will also increase due to the accumulating integrator state in the PI controller of the PLL, which is the LOS mechanism of WTG during the SCF.

*3.1.2 Discussion on asymmetric fault:*

To investigate the fault type influences towards the TSS, sequential circuit of OWF-MMC system is analysed. Suppose a SCF is imposed on the sending cable as shown in Fig. 6, and its positive sequence circuit is shown in Fig. 7, where the effects of negative/zero sequence is represented by an equivalent impedance $Z_{eq}$, the positive sequence voltage of MMC is $U_{mmc+}$ with maximum current limitation of $I_{lim}$, and $Z_{g1+}$ along with $Z_{g2+}$ denote the residual positive sequence impedance of the circuit, which is mostly the summation of the WTG and OWF transformer leakage impedance due to short cable lines. Then, four types of SCF are analysed below. According to the analysis in [8], $Z_{eq}$ is relatively small when single-phase and interphase SCF occurs, due to limited negative sequence current, and the maximum positive sequence voltage of MMC could be 0.57 and 0.5, respectively. Therefore, condition (6) is not likely to be violated because:

$$U_g \approx 0.5,\ I_{sd} < 1.5,\ Z_g \approx Z_{g1+} + Z_{g2+} \approx 0.2$$

On the other hand, if two-phase grounding fault occurs on the sending cable, then $Z_{eq}$ will be very low because the zero sequence impedance on the sending cable system is very small, due to the connection of the transformers. Besides, for three-phase SCF on the sending cable, the case is even worse for $Z_{eq}$ is almost zero. Therefore, if two-phase or three-phase grounding fault happens in the sending cable, the positive sequence voltage at the fault point $U_{f+}$ will be very close to zero because of the current limitation of the MMC station,

and consequently the current injection boundary (6) is most likely to be violated, because:

$$U_g \approx 0,\ I_{sd} < 1.5,\ Z_g \approx Z_{g1+}$$

Then, the frequency of PLL would diverge during these two kinds of SCF. Moreover, it is also indicated that the closer SCF to the MMC station, the faster the PLL frequency is climbing, for in this case $Z_g$ is the largest which makes $U_0$ much greater than $U_g \sin\delta$. In conclusion, the worst condition to lead to synchronization instability is the three-phase SCF on the sending cable near the MMC station.

*3.1.3 LOS of WTG after the fault clearance:*

Though the PLL frequency is diverging and the phase in not locked during the fault, it might either get resynchronized or suffer the LOS after the fault is cleared. Continuous low voltage caused by SCF is not realistic for the faulted cable will be quickly removed by the relay protection devices. The post-fault TSS make more sense and its mechanism could be explained by the classic equal-area criterion (EAC), thanks to the similarity between the PLL dynamics and the rotor movement of the synchronous generator. To understand this better, we substitute $\omega = x_{int} + k_p U_{sq}$ as a new state variable, and then equations (5) is reformed to [14]:

$$\begin{cases} \dot\delta = \omega_\delta \\ H_{pll}\dot\omega = T_m - T_e(\delta) - D_{pll}(\delta)\omega_\delta \end{cases} \quad (8)$$

where $T_m = U_0$ is equivalent to the mechanical torque, $T_e(\delta) = U_g \sin\delta$ the electric torque, $H_{pll} = (1 - k_p X_g I_{sd}/\omega_s)/k_i$ the inertia, and $D_{pll}(\delta) = k_p U_g \cos\delta/k_i - X_g I_{sd}/\omega_s$ the damping. To apply the EAC method, the damping $D(\delta_{pll})$ is neglected. Before the fault occurrence, the system is working at point $A$, where $l_1$ represents the electrical torque and $T_{m1}$ the mechanical torque. When SCF occurs, $U_g$ drops so that the electrical torque falls to $l_2$, which is lower than the mechanical torque, and thus the angle difference $\delta$ increases due to the dynamics of the PLL. The frequency accumulation is represented by $S_I$ during the fault. After a while, when the fault circuit is removed at $\delta_1$, $U_g$ restores to its normal value and then electrical torque changes back to $l_1$, whereas the mechanical torque jumps to $T_{m2}$ because the grid impedance changes due to the removal of one circuit. At this time, PLL angle difference $\delta$ would keep on increasing because the instantaneous PLL frequency is larger than the synchronous frequency, and the area before it reaches the critical point $B$ is referred to as the decelerating area. If the square $S_{II}$ is larger than that of $S_I$, then the frequency will recover before the critical point and the PLL will be able to return to point $C$ and lock on the grid voltage again; otherwise, it would not be able to get synchronized with the grid, indicating that the WTG will suffer the LOS. It could be seen that the fault clearing time is one key factor to determine whether the WTG would suffer the LOS after the fault clearance, because a longer FCT implies a larger accelerating area $S_I$ and a smaller decelerating area $S_{II}$, so that the LOS would be more likely to occur.

In summary, when two-phase or three-phase SCF occurs at the sending cable, the grid voltage would be very low so that (6) is almost violated. Then the PLL angle of the WTG begins to increase and the frequency begins to accumulate. For the post-fault system when the fault has been cleared, if the decelerating area is greater than the accelerating area, then the system WTG will be able to get synchronized again with the MMC according to the EAC, otherwise the LOS would probably happen and cause fatal consequence like power oscillation and overvoltage.

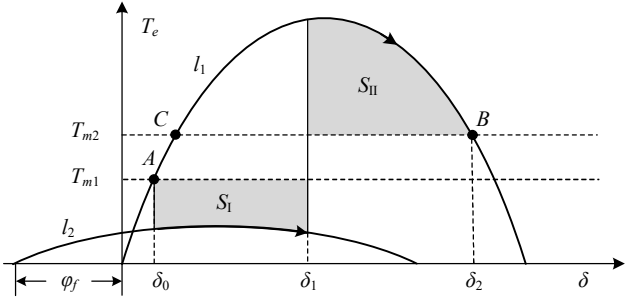

Fig. 8 LOS mechanism illustrated by the equal-area criterion (EAC).

*3.2 The Stability Index*

The EAC could be used to explain the basic principle of how LOS happens, but it is not rigorous for the TSS analysis, for the damping term $D_{pll}(\delta)$ is an indefinite function to $\delta$, whereas EAC requires a positive one. Therefore, EAC might cause error in FCT estimation [17, 22]. To this end, we refer to a rigorous function proposed by [17]:

$$V(\delta,x) = V_0 + \frac{1}{2}[x - h(\delta - \delta_s)]^2 - (1-\gamma h)(m\delta + \cos\delta), \quad (9)$$

where $V_0 = (1-\gamma h)(m\delta_s + \cos\delta_s)$, $\delta_s = \arcsin m$ is the angle difference at the stable equilibrium point (SEP), $m, \gamma, h, x$ are defined as the following substitutions:

$$\gamma = \frac{k_p \sqrt{U_g}}{\sqrt{k_i}}, \quad h = \frac{X_g I_{sd} \sqrt{k_i}}{\omega_g \sqrt{U_g}}, \quad m = \frac{U_0}{U_g}, \quad x = \frac{x_{int}}{\sqrt{k_i U_g}}, \quad (10)$$

This function serves as a local Lyapunov function to the PLL. It should be noted that all the parameters come from the post-fault system, because the aim is to estimate whether the system state is still inside the ROA of the post-fault system the moment after the fault is cleared. For simplicity, $h \approx 0$ is a reasonable approximation because $X_g I_{sd} \sqrt{k_i} \ll \omega_s$ is satisfied for the post-fault system. Then (9) is reduced to:

$$V(\delta,x) = \frac{1}{2}x^2 - (m\delta + \cos\delta) + (m\delta_s + \cos\delta_s). \quad (11)$$

And the value of the corresponding critical level set is:

$$V_{cr} = (m\delta_s + \cos\delta_s) - (m\delta_{cr} + \cos\delta_{cr}) \quad (12)$$

where $\delta_{cr} = \pi - \delta_s$ for $m > 0$ and $\delta_{cr} = -\pi - \delta_s$ for $m < 0$. If the LF value of the PLL is lower than the critical, i.e.:

$$V(\delta,x) < V_{cr}, \quad \delta \in [-\pi - \delta_s, \pi - \delta_s] \quad (13)$$

then the system state must be inside the ROA, indicating that the PLL would get synchronized after the fault is cleared; otherwise, the LOS might possibly happen. In this regard, we define the stability index:

$$\zeta = 1 - \frac{V(\delta,x)}{V_{cr}}, \quad \delta \in [-\pi - \delta_s, \pi - \delta_s], \quad (14)$$

where the angle difference $\delta$ could be estimated through the virtual orthogonal power method [23]. If the stability index is positive ($\zeta > 0$), then the system states are within the ROA and thus the PLL would be synchronization stable; otherwise, if the stability index is negative ($\zeta < 0$), then PLL is possible to suffer the LOS, yet it is not certain due to the damping effect of the proportional gain of the PLL [17]. This implies the stability index is conservative. Still, the result is acceptable because the conservativeness can always ensure that the counter-measurement is effectively taken, which is good for both converter safety and power system operation.

## 4 Simulation Verification

To show the feasibility of the proposed stability index in reflecting the TSS of the OWF-MMC system, a typical simulation scenario is designed, where a 1100MW MMC-HVDC system is connected with three OWFs #1~#3 of rated power 400MW, 300MW and 400MW, respectively (one could refer to Fig. 1 for the main connection), and the length of the sending cables are 12km, 6km and 3km from the OWF transformer platform to the offshore MMC station, where the double circuits are applied. There are 100 WTGs in #1 OWF, each of 4MW rated power, and the whole OWF is aggregated as one single WTG, as well as the feeder cable. Besides, 67×4.5MW and 100×4MW WTGs are configured in #2 and #3 OWFs, which are aggregated as well. The controlling structure of the OWF-MMC station is described in Section 2.2. Parameters of the transformers and cables are listed in Table 1.

The simulation is conducted on the PSCAD/EMTDC platform. SCF occurs at $t = 1.3$s on one of the circuits of the sending cable of the OWF #1 near the offshore MMC station, as is illustrated in Fig. 6. After a while, the faulted cable is removed by the CBs, and no block signal is applied to any of the converters in the whole duration. The configurations of the simulation cases are shown in Table 2, where six cases are analysed in comparison, including four methods for TSS improvement. It should be mentioned that if the LVRT control is enabled in the WTG, it means that the q-axis current

TABLE 1 Transformer and Cable Parameters.

| Description | Voltage | Parameter |
|---|---|---|
| WTG Transformer | 0.69/35kV | $S_n = 4.5\text{MVA}, U_k\% = 7$ |
| OWF Transformer | 35/220kV | $S_n = 480\text{MVA}, U_k\% = 10.5$ |
| Aggregated Feeder Cable | 35kV | $R = 0.038\Omega, X = 0.06\Omega, B = 172\Omega$ |
| Sending Cable | 220 kV | $r = 0.02\Omega/\text{km}, x = 0.4\Omega/\text{km}, b = 0.16\text{M}\Omega/\text{km}$ |

TABLE 2 Configurations of the Simulation Cases.

| Case Number | Fault Type | PI Params. of PLL | Fault Clearing Time | OWF Power | LVRT of WTG |
|---|---|---|---|---|---|
| Case 1 | ABC→G | (40, 1600) | 200ms | 1.0 pu | disable |
| Case 2 | A→G | (40, 1600) | 200ms | 1.0 pu | disable |
| Case 3 | ABC→G | (30, 900) | 200ms | 1.0 pu | disable |
| Case 4 | ABC→G | (40, 1600) | 100ms | 1.0 pu | disable |
| Case 5 | ABC→G | (40, 1600) | 200ms | 0.3 pu | disable |
| Case 6 | ABC→G | (40, 1600) | 200ms | 1.0 pu | enable |

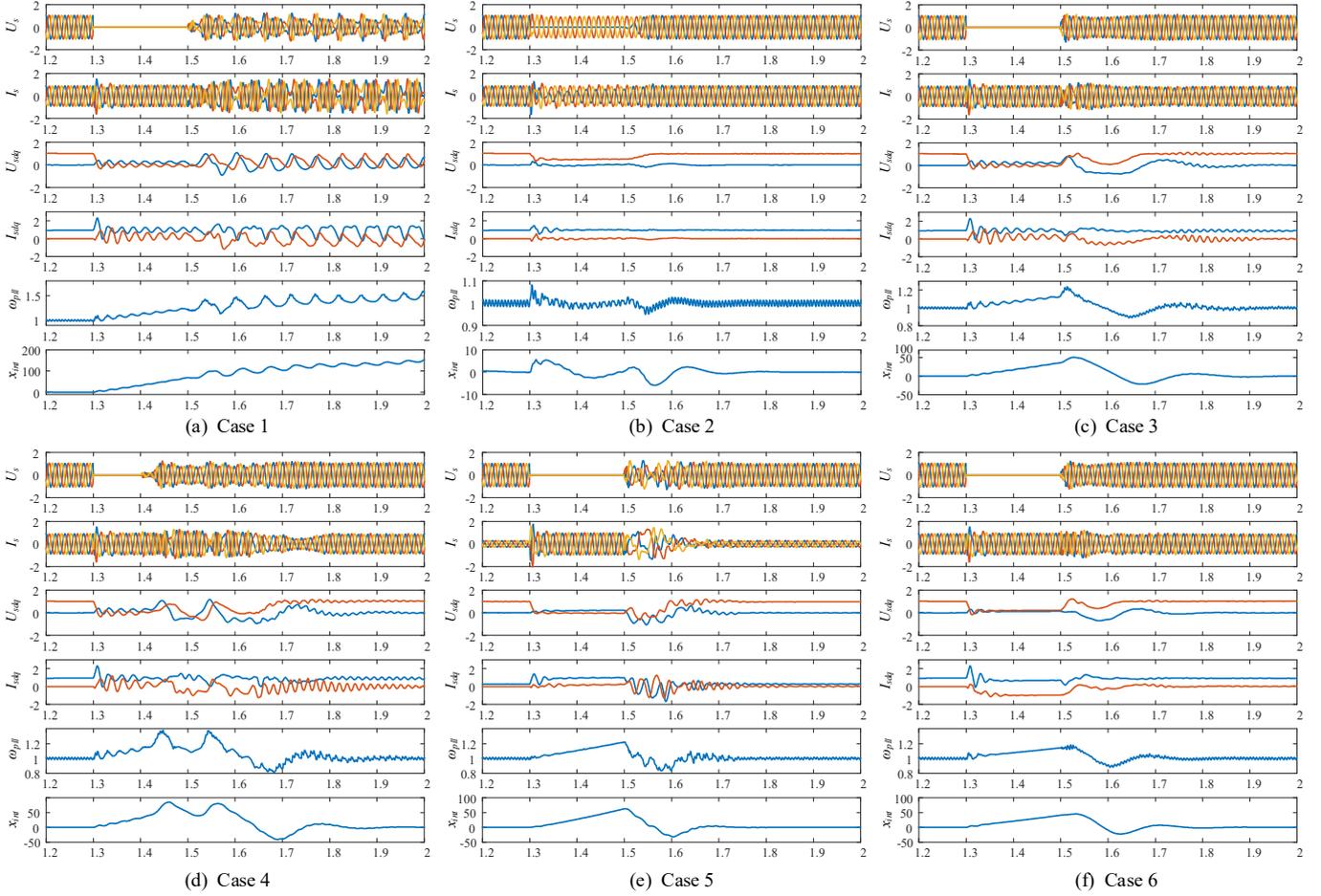

Fig. 9 Simulation result of different cases ($U_s$ [p.u.] denotes the voltage of the offshore MMC station, $I_s$ [p.u.] the current of the offshore MMC station, $U_{sdq}$ [p.u.] the $dq$-axis voltage of the aggregated OWF #1, $I_{sdq}$ [p.u.] the $dq$-axis current of the aggregated OWF #1, $\omega_{pll}$ [rad/s] the PLL frequency of the aggregated OWF #1, $x_{int}$ [rad/s] the integrator state of the PLL of the aggregated OWF #1)

reference is computed by (1), and thus the $d$-axis current is forced to decrease due to the maximum current limitation, which is set to 1.2 pu in the simulation.

The simulation results are shown in Fig. 9. It is shown in Fig. 9 (a) that the OWF-MMC system could suffer the LOS if three phase SCF occurs. During the fault ($1.3s < t < 1.5s$), the frequency of the PLL is increasing, leading to the temporary LOS of the WTG. After the fault is cleared, the frequency of PLL does not restore back, but continuously increases and couldn't lock the grid voltage angle again, leading to the oscillation of the $dq$-axis voltage in the control. Moreover, the voltage and current profile of the offshore AC grid is distorted due to the LOS of the WTG. In comparison, as is shown in Fig. 9 (b), if the fault type is the single phase SCF, then the WTG would not suffer the LOS even during the fault, for the positive sequence voltage is still relatively large. After the faulted cable is cleared, the voltage and current profile could restore soon. Next, it is demonstrated in Fig. 9 (c)-(f) that the TSS of the OWF could be improved by: 1) reducing the PLL bandwidth, 2) a faster FCT, 3) lower active power generation, 4) enabling the LVRT control. For case 3, the bandwidth of PLL is 30rad/s, which is lower than 40rad/s for that of the case 1. Therefore, the increasing rate of the PLL frequency is slower than that of the case 1. The angle difference $\delta$ does not reach the critical value after the clearance of the fault, and hence the PLL is able to get resynchronized with the grid, as is shown in Fig. 9 (c). Similarly, if the fault is cleared faster, then the PLL frequency does not deviate too much so that it is also able to restore after the fault is removed, as is shown in Fig. 9 (d). Apart from either reducing the PLL bandwidth or shortening the FCT, to reduce the $d$-axis current is also an effective way to improve the TSS. As is implied in case 5, the $d$-axis current of the post fault system is much lower than that of the case 1, indicating that the decelerating area could be larger, as is illustrated in Fig. 8, and thus the TSS could be improved, as is shown in Fig. 9 (e). In case 6, the $d$-axis current is forced to reduce due to the injection of $q$-axis current, leading to a smaller accelerating area, and thus the TSS could be improved as well, as is shown in Fig. 9 (f).

Moreover, to compare these cases quantitatively, their corresponding stability index is computed, as is shown in Fig. 10. It could be firstly seen that the index is decreasingly negative for case 1 while eventually positive for case 2~6, exhibiting the same stability result as the time domain simulation in Fig. 9. During the SCF, the index of case 2 is almost constant 1, implying that the WTG is always stable for single phase SCF even during the fault. Then, from the stability index curve of the other cases, it can be concluded that case 6 has the highest stability margin, then followed by case 5, 3, 4 and finally 1, indicating that the LVRT control improves the TSS of the WTG the most, whereas to reduce the PLL bandwidth or the FCT might be less effective. Besides, injecting $q$-axis reactive current also helps to improve the grid voltage. Therefore, the LVRT control is recommended in the offshore WTGs for the good of both TSS and voltage supporting.

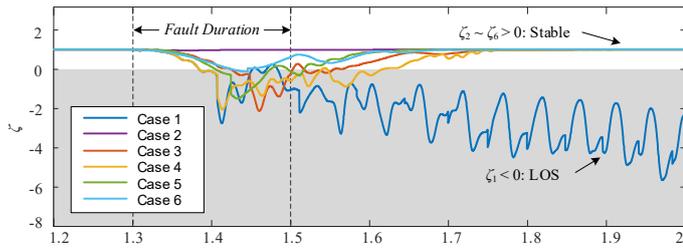

Fig. 10 The stability index of each case.

## 5 Conclusion

In this paper, the mechanism of the LOS of the OWF-MMC system is revealed, considering FRT control, fault type, and fault clearing time. A stability index is proposed based on one previous Lyapunov function, which could be used to estimate the TSS and quantify the stability margin. It is shown in the simulation that the offshore WTG could suffer the LOS after suffering three-phase SCF on the sending cable, and the stability can be improved by several prevailing methods. The performance of these methods is evaluated by the proposed stability index, and it shows that the most effective way to enhance the TSS is by injecting $q$-axis reactive current, compared with the others. More detailed and quantitative analysis will be shown in the future works.

## References

[1] J. Lyu, X. Cai, and M. Molinas, "Optimal Design of Controller Parameters for Improving the Stability of MMC-HVDC for Wind Farm Integration," *IEEE Journal of Emerging and Selected Topics in Power Electronics,* vol. 6, no. 1, pp. 40-53, 2018, doi: 10.1109/jestpe.2017.2759096.

[2] H. Zong, C. Zhang, J. Lyu, X. Cai, M. Molinas, and F. Rao, "Generalized MIMO Sequence Impedance Modeling and Stability Analysis of MMC-HVDC with Wind Farm Considering Frequency Couplings," *IEEE Access,* vol. PP, pp. 1-1, 03/16 2020, doi: 10.1109/ACCESS.2020.2981177.

[3] L. Xu, L. Yao, and C. Sasse, "Grid Integration of Large DFIG-Based Wind Farms Using VSC Transmission," *IEEE Transactions on Power Systems,* vol. 22, no. 3, pp. 976-984, 2007, doi: 10.1109/tpwrs.2007.901306.

[4] B. Liu, J. Xu, R. E. Torres-Olguin, and T. Undeland, "Faults mitigation control design for grid integration of offshore wind farms and oil & gas installations using VSC HVDC," in *SPEEDAM 2010,* 14-16 June 2010 2010, pp. 792-797, doi: 10.1109/SPEEDAM.2010.5542252.

[5] U. Karaagac, J. Mahseredjian, L. Cai, and H. Saad, "Offshore Wind Farm Modeling Accuracy and Efficiency in MMC-Based Multiterminal HVDC Connection," *Ieee T Power Deliver,* vol. 32, no. 2, pp. 617-627, 2017, doi: 10.1109/tpwrd.2016.2522562.

[6] S. K. Chaudhary, R. Teodorescu, P. Rodriguez, P. C. Kjaer, and A. M. Gole, "Negative Sequence Current Control in Wind Power Plants With VSC-HVDC Connection," *Ieee T Sustain Energ,* vol. 3, no. 3, pp. 535-544, 2012, doi: 10.1109/tste.2012.2191581.

[7] J. Freytes, J. Li, G. De-Preville, and M. Thouvenin, "Grid-Forming Control with Current Limitation for MMC under Unbalanced Fault Ride-Through," *Ieee T Power Deliver,* pp. 1-1, 2021, doi: 10.1109/tpwrd.2021.3053148.

[8] L. Shi, G. P. Adam, R. Li, and L. Xu, "Control of Offshore MMC During Asymmetric Offshore AC Faults for Wind Power Transmission," *IEEE Journal of Emerging and Selected Topics in Power Electronics,* vol. 8, no. 2, pp. 1074-1083, 2020, doi: 10.1109/jestpe.2019.2930399.

[9] Y. Li et al., "Over-Voltage Suppression Methods for the MMC-VSC-HVDC Wind Farm Integration System," *IEEE Transactions on Circuits and Systems II: Express Briefs,* vol. 67, no. 2, pp. 355-359, 2020, doi: 10.1109/TCSII.2019.2911652.

[10] C. Liu, F. Deng, Q. Heng, X. Cai, R. Zhu, and M. Liserre, "Crossing Thyristor Branches Based Hybrid Modular Multilevel Converters for DC Line Faults," *IEEE Transactions on Industrial Electronics,* 2020.

[11] Y. Li et al., "PLL Synchronization Stability Analysis of MMC-Connected Wind Farms under High-Impedance AC Faults," *IEEE Transactions on Power Systems,* vol. 36, no. 3, pp. 2251-2261, 2021, doi: 10.1109/TPWRS.2020.3025917.

[12] D. Dong, B. Wen, D. Boroyevich, P. Mattavelli, and Y. Xue, "Analysis of phase-locked loop low-frequency stability in three-phase grid-connected power converters considering impedance interactions," *IEEE Transactions on Industrial Electronics,* vol. 62, no. 1, pp. 310-321, 2015, doi: 10.1109/TIE.2014.2334665.

[13] X. He, H. Geng, R. Li, and B. C. Pal, "Transient Stability Analysis and Enhancement of Renewable Energy Conversion System During LVRT," *Ieee T Sustain Energ,* vol. 11, no. 3, pp. 1612-1623, 2020, doi: 10.1109/TSTE.2019.2932613.

[14] C. Zhang, X. Cai, A. Rygg, and M. Molinas, "Modeling and analysis of grid-synchronizing stability of a Type-IV wind turbine under grid faults," *International Journal of Electrical Power and Energy Systems,* vol. 117, no. July 2018, p. 105544, 2020, doi: 10.1016/j.ijepes.2019.105544.

[15] X. Fu et al., "Large-Signal Stability of Grid-Forming and Grid-Following Controls in Voltage Source Converter: A Comparative Study," *Ieee T Power Electr,* vol. 36, no. 7, pp. 7832-7840, Jul 2021, doi: 10.1109/tpel.2020.3047480.

[16] Q. Hu, L. Fu, F. Ma, and F. Ji, "Large Signal Synchronizing Instability of PLL-Based VSC Connected to Weak AC Grid," *IEEE Transactions on Power Systems,* vol. 34, no. 4, pp. 3220-3229, 2019, doi: 10.1109/TPWRS.2019.2892224.

[17] Y. Zhang, C. Zhang, and X. Cai, "Large-Signal Grid-synchronization Stability Analysis of PLL-based VSCs Using Lyapunovs Direct Method," *IEEE Transactions on Power Systems,* pp. 1-1, 2021, doi: 10.1109/tpwrs.2021.3089025.

[18] P. Rodríguez, A. Luna, R. S. Muñoz-Aguilar, I. Etxeberria-Otadui, R. Teodorescu, and F. Blaabjerg, "A stationary reference frame grid synchronization system for three-phase grid-connected power converters under adverse grid conditions," *Ieee T Power Electr,* vol. 27, no. 1, pp. 99-112, 2012, doi: 10.1109/TPEL.2011.2159242.

[19] M. Graungaard Taul, X. Wang, P. Davari, and F. Blaabjerg, "Current Reference Generation Based on Next-Generation Grid Code Requirements of Grid-Tied Converters During Asymmetrical Faults," *IEEE Journal of Emerging and Selected Topics in Power Electronics,* vol. 8, no. 4, pp. 3784-3797, 2020, doi: 10.1109/jestpe.2019.2931726.

[20] S. K. Chung, "A phase tracking system for three phase utility interface inverters," *Ieee T Power Electr,* vol. 15, no. 3, pp. 431-438, 2000, doi: 10.1109/63.844502.

[21] E. Muljadi, S. Pasupulati, A. Ellis, and D. Kosterov, "Method of equivalencing for a large wind power plant with multiple turbine representation," in *2008 IEEE Power and Energy Society General Meeting - Conversion and Delivery of Electrical Energy in the 21st Century,* 20-24 July 2008 2008, pp. 1-9, doi: 10.1109/PES.2008.4596055.

[22] C. Zhang, M. Molinas, Z. Li, and X. Cai, "Synchronizing Stability Analysis and Region of Attraction Estimation of Grid-Feeding VSCs Using Sum-of-Squares Programming," *Frontiers in Energy Research,* vol. 8, no. April, pp. 1-12, 2020, doi: 10.3389/fenrg.2020.00056.

[23] R. Sun, J. Ma, W. Yang, S. Wang, and T. Liu, "Transient Synchronization Stability Control for LVRT with Power Angle Estimation," *Ieee T Power Electr,* vol. PP, pp. 1-1, 04/01 2021, doi: 10.1109/TPEL.2021.3070380.